# Electron-phonon coupling in semiconductors at high electronic temperatures

N. Medvedev[1,2,*]

1. Institute of Physics, Czech Academy of Sciences, Na Slovance 2, 182 21 Prague 8, Czech Republic

2. Institute of Plasma Physics, Czech Academy of Sciences, Za Slovankou 3, 182 00 Prague 8, Czech Republic

## Abstract

A nonperturbative dynamical coupling approach based on tight-binding molecular dynamics is used to evaluate the electron-ion (electron-phonon) coupling parameter in irradiated semiconductors as a function of the electronic temperature up to ~25,000 K. The method accounts for arbitrary electronic distribution function *via* the Boltzmann equation, enabling a comparative analysis of various models: fully equilibrium electronic distribution, band-resolved local equilibria (distinct temperatures and chemical potential of electrons in the valence and the conduction band), and a full nonequilibrium distribution. It is demonstrated that the nonequilibrium produces the electron-phonon coupling parameter different by at most ~35% from its equilibrium counterpart for identical deposited energy density, allowing to use the coupling parameter as a function of the single electronic equivalent (or kinetic) temperature. The following 14 semiconductors are studied here – group IV: Si, Ge, SiC; group III-V: AlAs, AlP, GaP, GaAs, GaSb; oxides: ZnO, $TiO_2$, $Cu_2O$; layered $PbI_2$; ZnS and $B_4C$.

## I. Introduction

Treatments of semiconductors with laser irradiation are used in various technological applications, such as the production of nanochips, designing their properties, and micromachining [1–6], along with basic research [7,8]. Irradiation of semiconductors with ultrafast powerful pulses potentially enables atomic-scale control of the material modifications on the surface [6,9].

The most commonly used theory of the material response to laser irradiation is the two-temperature model (TTM) [10–12], which describes the electronic and the atomic systems of the material as separate systems each in the local thermal equilibrium state with distinct temperatures. Such a description is most suitable for metals but poses a number of challenges for bandgap materials (semiconductors and insulators) [12]. In these kinds of materials, the electronic ensemble is separated into two parts: valence-band and conduction-band electrons, each possessing different properties. This complicates the description of the electronic ensemble with a single thermodynamical equation used in the TTM.

More advanced models exist to describe insulators' and semiconductors' response to irradiation, which treat the valence- and the conduction-band electrons separately: fully nonequilibrium simulation techniques (e.g., based on the Boltzmann equation [13,14]), or models assuming separate local equilibria in the valence and the conduction band (e.g., the so-called nTTM model [15,16], or the three-temperature model, 3TM [17]).

---

[*] Corresponding author's email: nikita.medvedev@fzu.cz

In all such models, the key parameter governing the energy exchange between the electronic and the atomic ensemble is the electron-phonon coupling parameter. The standard theory of calculation of the coupling parameter – the Eliashberg spectral function formalism [18] – cannot straightforwardly be applied to bandgap materials but requires a careful treatment of the density of states around the Fermi level. Often, the coupling parameter is used as a fitting parameter in attempts to reproduce the observable material damage in a simulation [19]. Thus, methods evaluating the electron-phonon coupling parameter at high electronic temperatures are in high demand. Obtaining the coupling parameters for commonly used semiconductors should enable the application of models without adjustable parameters to better understand and control the processes involved in ultrafast irradiation. To do so, the problem of the electronic nonequilibrium induced by the presence of the bandgap should be considered. Namely, the influence of the possible nonequilibrium electronic distribution function on the coupling parameter should be elucidated.

Here, a recently developed method based on a hybrid model combining the Boltzmann equation with the tight-binding molecular dynamics simulation is used to nonperturbatively calculate the electron-phonon coupling parameter in a wide range of semiconductors [20,21]. The dynamical coupling method treats the nonadiabatic response of the electronic populations to arbitrary atomic displacements, thereby enabling the evaluation of the coupling between the two systems. Most importantly, the methodology allows to evaluate the importance of the nonequilibrium electronic distribution function on the coupling parameter.

## II. Model

For evaluation of the electron-ion (electron-phonon) coupling parameter, XTANT-3 code is used [22]. It combines (a) a Monte Carlo (MC) model for photoabsorption and nonequilibrium kinetics of high-energy electrons; (b) Boltzmann equation (BE) describing the low-energy electron (populating the valence band and the bottom of the conduction band) evolution on the transient band structure; (c) a transferable tight binding (TB) for a description of the transient band structure and the interatomic potential; and (d) molecular dynamics (MD) simulation for propagation of atomic trajectories. It has been previously applied for the evaluation of the coupling parameters in metals in the case of electronic equilibrium [20]. To apply it to semiconductors, it has been extended to include electronic nonequilibrium and various limiting cases of possible thermalization. To this end, the evolution of the electronic distribution function is modeled with the Boltzmann equation including the collision integrals [21]:

$$\frac{df_e(\varepsilon_i, t)}{dt} = I_{e-e} + I_{e-i} + I_{MC}. \qquad (1)$$

The distribution function, $f_e$, describes the fractional electron populations on the transient electronic energy levels (molecular orbitals, $\varepsilon_i = \langle \psi_i(t)|H|\psi_i(t)\rangle$, with $H$ being the TB Hamiltonian dependent on the positions of all the atoms in the simulation box); $I_{e-e}$ is the electron-electron scattering integral; $I_{e-i}$ is the electron-ion (electron-phonon) collision integral; and $I_{MC}$ is the source term responsible for the photoabsorption and interaction with the high-energy electrons, if any [21].

The electron-electron thermalization is described with the help of the relaxation time approximation [21].:

$$I_{e-e} = -\frac{f_e(\varepsilon_i, t) - f_{eq}(\varepsilon_i, \mu, T_e, t)}{\tau_{e-e}}. \qquad (2)$$

with $\tau_{e-e}$ the characteristic global electron relaxation time; $f_{eq}(\varepsilon_i, \mu, T_e, t)$ is the equivalent equilibrium Fermi-Dirac distribution with the same total number of (low-energy) electrons ($n_e$) and energy content ($E_e$) as in the transient nonequilibrium distribution:

$$\begin{cases} n_e = \sum f_e(\varepsilon_i, t) = \sum f_{eq}(\varepsilon_i, \mu, T_e, t) \\ E_e = \sum \varepsilon_i f_e(\varepsilon_i, t) = \sum \varepsilon_i f_{eq}(\varepsilon_i, \mu, T_e, t) \end{cases}. \qquad (3)$$

Eqs. (3) define the equivalent electronic temperature (also called the kinetic temperature, $T_e$ [23]) and the equivalent chemical potential ($\mu$) [21]. The same Eqs. (2-3) can be used separately for the electrons in the valence and the conduction band, with their own thermalization times, thereby allowing for separate partial thermalizations of the electronic fractions.

Note that the choice of the thermalization times automatically produces various limiting cases: setting the global thermalization time $\tau_{e-e} \to 0$ makes the instantaneously globally thermalized electronic ensemble, thus reducing the model to the two-temperature-based (TTM) TBMD; setting separate the valence- and the conduction-band thermalization times $\tau_{e-e,\{v,c\}} \to 0$ creates two electronic ensembles thermalized at their own chemical potentials and temperatures restoring the three-temperature model, 3TM; using finite thermalization times (either global or band-resolved partial) produces a nonequilibrium electronic ensemble. This flexibility allows us to study various cases and the influence of the model approximation for the electronic distribution on the coupling parameter. Further on, the electronic temperature wherever used is the equivalent (or kinetic) temperature defined by Eqs.(3).

The electron-phonon coupling is extracted from the Boltzmann collision integral [20]:

$$I_{e-i}^{ij} = w_{ij} \begin{cases} f_e(E_j)(2 - f_e(E_i))e^{-E_{ij}/T_a} - f_e(E_i)(2 - f_e(E_j)), \text{ for } i > j \\ f_e(E_j)(2 - f_e(E_i)) - f_e(E_i)(2 - f_e(E_j))e^{-E_{ji}/T_a}, \text{ for } i < j. \end{cases} \qquad (4)$$

here $E_{ij} = E_i - E_j$ is the difference between the energies of the two levels; $T_a$ is the atomic temperature in the Maxwellian distribution; and $w_{ij}$ is the rate of electron transitions triggered by atomic motion approximated in the dynamical coupling formalism as [20]:

$$w_{ij} \approx \frac{4e}{\hbar \delta t^2} \sum_{\alpha,\beta} |c_{i,\alpha}(t) c_{j,\beta}(t_0) S_{i,j}|^2. \qquad (5)$$

where the wave functions, $\psi(t)$, are calculated on two consecutive steps in the molecular dynamics simulation: $t_0$ and $t = t_0 + \delta t$, to obtain $c_{i,\alpha}$, the coefficients in the linear combination of atomic orbitals basis set within the TB Hamiltonian ($\psi_i = \sum_\alpha c_{i,\alpha} \varphi_\alpha$); $S_{\alpha,\beta}$ is the TB overlap matrix; $e$ is the electron charge, and $\hbar$ is Planck's constant.

The electron-phonon coupling parameter is then calculated from the electron-ion collision integral [20]:

$$G(T_e, T_a) = \frac{1}{V(T_e - T_a)} \sum_{i,j} E_i I_{e-a}^{ij}. \quad (6)$$

Here *V* is the volume of the simulation box; again, $T_e$ denotes the equivalent (kinetic) electronic temperature, thereby defining the coupling parameter for nonequilibrium electronic distributions in a generalized way allowing for a meaningful comparison between equilibrium and various nonequilibrium scenarios. In the equilibrium case, it naturally reduces to the thermodynamic temperature, restoring the standard definition.

Following the previous work on metals [20], for each material, ten XTANT-3 simulations are run with different initial conditions and parameters of irradiation (various pulse durations and deposited doses), to obtain reliable averaged results reducing the influence of statistical fluctuations [24]. The methodology follows the kinetics in the simulation box in time while increasing the electronic temperature, thus extracting the dependency of the dynamically calculated coupling on the kinetic electronic temperature.

Depending on the duration of the increase of the electronic temperature, we may include or exclude the nonthermal effects: atomic acceleration due to the changes in the interatomic potential caused by the electronic excitation [25]. The most famous effect is the so-called nonthermal melting, in which the atomic lattice destabilizes directly due to the electronic excitation, before any significant electron-phonon coupling [26,27]. This is a distinct channel from the electron-phonon coupling and thus should be treated separately from the coupling parameter, as will be discussed below [25].

In all the simulations, an NVE (microcanonical) ensemble is used with periodic boundary conditions and an MD timestep of 0.1 fs with the Martyna-Tuckerman 4[th]-order algorithm [28].

Additionally, the electronic heat capacity is calculated as follows:

$$C_e(T_e) = \frac{1}{V} \sum_i \frac{\partial f_e(E_i)}{\partial T_e} (E_i - \mu(T_e)), \quad (7)$$

where for the calculation of the derivative $\partial f_e(E_i)/\partial T_e$, the derivative of the equivalent electronic chemical potential by the electronic kinetic temperature $\partial \mu(T_e)/\partial T_e$ is calculated numerically [29].

### III. Results

### III.1. Effect of nonequilibrium electron distribution on the electron-phonon coupling in semiconductors

Let us start by analyzing the influence of the possible nonequilibrium state on the electron-phonon coupling parameter on the example of GaAs. A set of simulations were performed: a full nonequilibrium simulation with a finite thermalization time of 10 fs and 100 fs, a full equilibrium one (full instantaneous thermalization), and a few simulations with band-resolved partial thermalization for the cases of irradiation with various photon energies (the three-temperature model). The last point is necessary for the demonstration of possible effects of the particular excitation scenarios since different photon energies create excited electron ensembles with different energy densities in the separate bands: initially, a single photo-electron is emitted with energy dependent on the photon energy, which then may thermalize inside the conduction band to create different states of the excited electron ensembles.

Examples of the electronic distribution functions for the same system are displayed in Figure 1. In this figure, GaAs was excited with the laser pulse of 10 fs FWHM duration and 3 eV photon energy and the deposited dose of 4 eV/atom. The electronic distribution may be noticeably different within different approximations; however, it is *a priori* unclear how this will affect thermodynamical parameters such as the electron-phonon coupling.

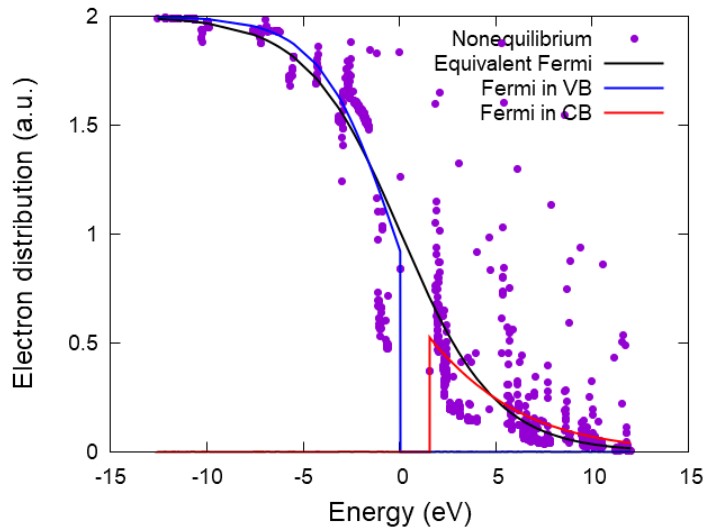

*Figure 1. Example of various approximations for the electronic distribution function in GaAs 10 fs after deposition of the dose of 4 eV/atom with the photon energy of 3 eV. The full nonequilibrium distribution is shown with circles, the separate equivalent Fermi distributions for the valence-band (VB) and conduction-band (CB) fractions of electrons, and the full equilibrium equivalent Fermi distributions are shown.*

The electron-phonon coupling parameter as a function of the equivalent (kinetic) temperature for this set of simulations is shown in Figure 2. One can see that nonequilibrium affects the coupling parameter only to a relatively small degree: the full nonequilibrium distribution deviates from the equilibrium value maximum by ~35% (below the equilibrium value), and by about the same amount for 3 eV and 6 eV photons within the 3TM approximation (separate thermalization of the valence and the conduction band; the curves above the equilibrium one). For other studied photon energies, the deviation is even smaller.

It seems to be a typical case that the nonequilibrium electron distribution produces lower coupling than the equilibrium one because the distribution, even though it has certain spikes, also has many almost unperturbed parts in between them which cannot participate efficiently in the energy exchange with the atoms/phonons [21,30]. In contrast, partial thermalization may have higher coupling than the full equilibrium one, because, for the same amount of the energy content in the electronic ensemble, either of the bands may have a significantly higher temperature than the equivalent fully-equilibrated one, thus increasing the coupling (e.g., in Figure 1, the fully-equilibrium equivalent temperature is $T_e = 28500$ K, whereas the partial equivalent temperatures are in the valence band $T_{e,v} = 22500$ K and in the conduction band $T_{e,c} = 42600$ K). In most cases, however, the deviation is rather small.

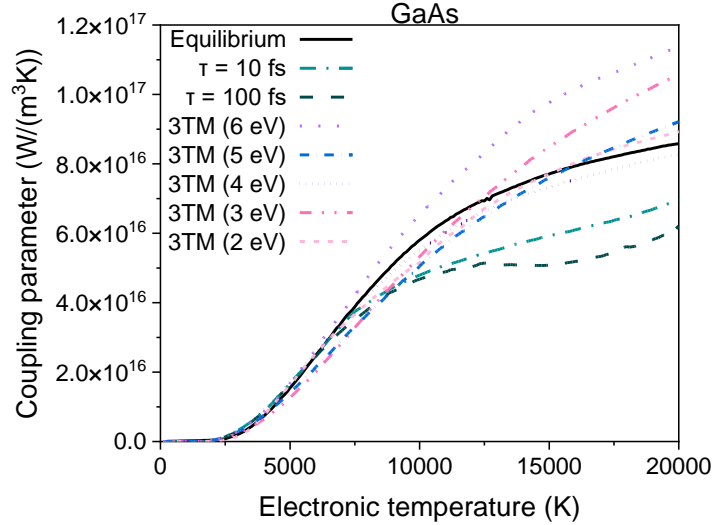

*Figure 2. Electron-phonon coupling in GaAs in various approximations: fully equilibrium, nonequilibrium with the thermalization times of 10 fs and 100 fs, and rand-resolved thermalizations (3TM) for the cases of various photon energies of the irradiating laser pulse. In all cases, the equivalent (kinetic) temperature (Eqs. (3)) is used for the X-axis.*

The physical scenario, realized in experiments, is typically starting with a full nonequilibrium distribution, which then partially thermalizes separately within the bands typically on femtosecond timescales. After that, the interband thermalization takes place, whose speed depends on the excitation level – the more electrons are excited, the faster they thermalize via the impact ionization and three-body recombination processes. It is expected to take place on the scale of a few tens to a few hundreds of femtoseconds. During those not-fully-equilibrated stages, the coupling parameter may deviate from the equilibrium one, but only within the range of some 35%. Thus, in most of the cases, the equilibrium value of the electron-phonon coupling parameter may be reliably used in models, as the function of the equivalent (kinetic) electron temperature.

We will proceed with the calculation of the electron-phonon coupling assuming full electronic equilibration in the sections below.

### III.2. Nonthermal effects on the electron-phonon coupling in semiconductors

Apart from the nonequilibrium effects, it is also important to keep in mind that covalent materials have another channel of the electron-ion energy exchange: nonthermal changes in the interatomic potential, which may accelerate the atoms [25]. This effect is most prominent at the deposited doses above the nonthermal damage threshold. At below (but close to) the threshold doses, this effect is known as the displacive excitation of coherent phonons; at above the threshold doses, it is known as nonthermal melting. This effect takes place even without any contribution from the electron-phonon energy exchange – it is an adiabatic effect. However, nonadiabatic effects can create synergy with the nonthermal ones, as discussed e.g. in Ref. [25]. Nonthermal acceleration of atoms increases the atomic temperature extremely fast. The electron-phonon coupling parameter nearly linearly depends on the atomic temperature [20] – thus, nonthermal acceleration also increases the coupling parameter. In turn, it heats the atoms even more, and this self-amplifying process leads to extremely fast damage in the covalent material and equilibration of the electronic and atomic temperatures [25].

An example of the influence of the nonthermal atomic acceleration on the electron-phonon coupling in silicon is shown in Figure 3. The coupling parameter, affected by ultrafast nonthermal atomic acceleration (reported in Ref. [20]) shows a dramatic increase around the electronic temperatures associated with the nonthermal melting. If this effect is excluded, the trend of the increase of the coupling parameter does not change.

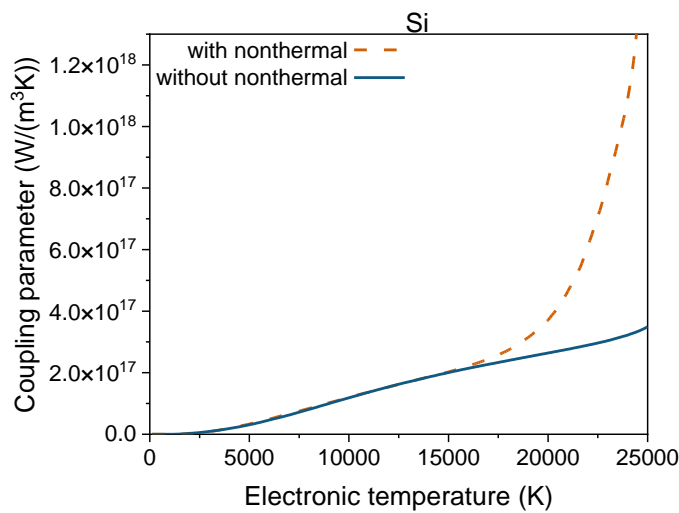

*Figure 3. Electron-phonon coupling in Si for the cases of including the effects of nonthermal atomic acceleration (as reported in Ref. [20]), and excluding it.*

These nonthermal effects are naturally occurring in XTANT-3, as they would in any nonadiabatic *ab-initio* simulation. They may be accounted for in other models in various ways: e.g., by directly implementing the coupling parameter with the increase at high electronic temperatures, or (more appropriately) by taking into account the dependence of the coupling parameter on the atomic temperature plus some additional terms accounting for the nonthermal acceleration upon reaching certain electronic temperature (for instance, see recently implemented classical MD simulation combined with the MC tracing transport of electrons and valence holes in [31]).

In all further simulations, we show the pure electron-phonon coupling, without the additional effect of the nonthermal atomic acceleration: we ensure that the coupling parameters are all calculated at room atomic temperature.

## III.2. Group IV semiconductors

Having established that the nonequilibrium effects may often be neglected, we may proceed now with evaluation of the coupling parameters in various classes of semiconductors under the approximation of the local thermal equilibrium in the electronic ensemble (the two-temperature state). The following parameters are used to calculate the coupling parameters and electronic heat capacities: for Si and Ge, the NRL tight-binding parameterization is used with 216 atoms in the simulation box (a comparison with other available TB parameterizations is discussed in Appendix) [32,33]; for SiC (in the hexagonal P6$_3$mc state [34]), the matsci-0-3 DFTB parameterization is used with 192 atoms in the simulation box [35].

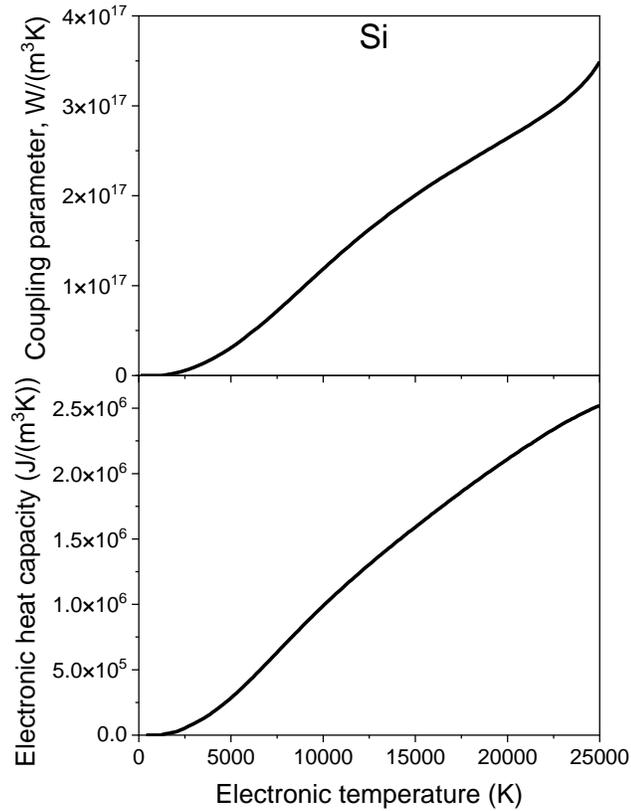

*Figure 4. (Top panel) Electron-phonon coupling, and (bottom panel) electronic heat capacity in Si.*

Figure 4 shows the XTANT-3 calculated parameters in Si. Nearly linear dependencies on the electronic temperature are seen in both, the coupling parameter and the electronic heat capacity. As was mentioned in Ref. [20], the coupling parameter is near zero up to the electronic temperature of ~2000 K – due to the presence of the band gap (~1.17 eV in cold Si) at lower temperatures, only exponentially small fraction of electrons is in the conduction band and correspondingly small number of holes in the valence band. A noticeable increase in the coupling takes place only when the electronic temperature becomes sufficient to promote a non-negligible number of electrons across the gap. Note that the nonthermal effects were excluded here, which may induce bandgap collapse at electronic temperatures ~17000 K [36].

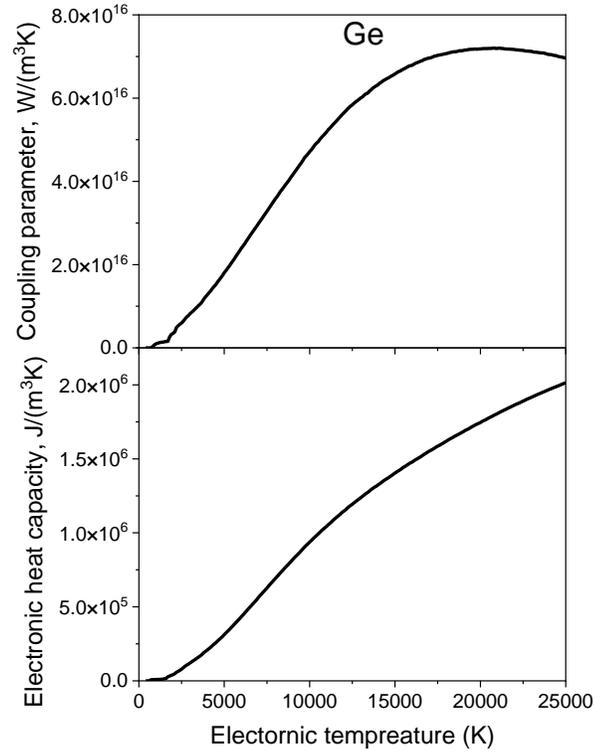

*Figure 5. (Top panel) Electron-phonon coupling, and (bottom panel) electronic heat capacity in Ge.*

A similar situation is in Ge, shown in Figure 5, but the absolute values of the electron-phonon coupling parameter are smaller than those in Si. This is in line with the trend of decreasing the coupling parameter with an increase in the atomic mass, discussed in Ref. [20]. The onset of the increase in both, the coupling parameter and electronic heat capacity, is earlier than in Si due to a smaller bandgap in Ge.

The coupling in SiC, presented in Figure 6, shows higher values, also in line with the trend: as carbon atoms are much lighter than Ge and Si, their faster motion allows for a stronger coupling with the electronic system.

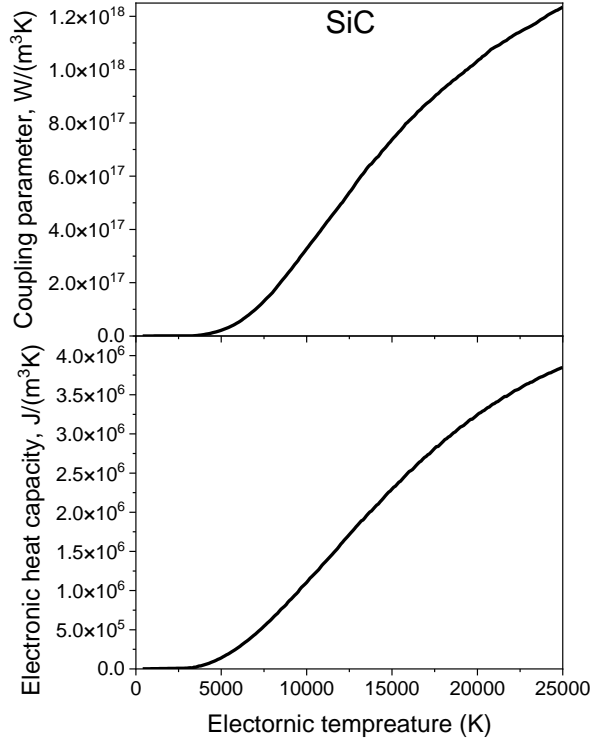

*Figure 6. (Top panel) Electron-phonon coupling, and (bottom panel) electronic heat capacity in SiC.*

### III.3. Group III-V semiconductors

For group III-V semiconductors (AlAs, AlP, GaAs, GaP, and GaSb) modeling, Molteni et al.'s TB parameterization was used [37], 216 atoms in the supercell in the zinc blende structure in each case. The AlAs calculated parameters are shown in Figure 7; for AlP in Figure 8; GaAs in Figure 9; GaP in Figure 10, and GaSb in Figure 11.

Among the materials in these group, the highest coupling is in the lightest elements: AlP, reaching $G\sim4\times10^{17}$ W/(m$^3$K) at the electronic temperature of $T_e\sim25000$ K; the smallest coupling, respectively, is in GaSb, reaching only $G\sim6\times10^{16}$ W/(m$^3$K).

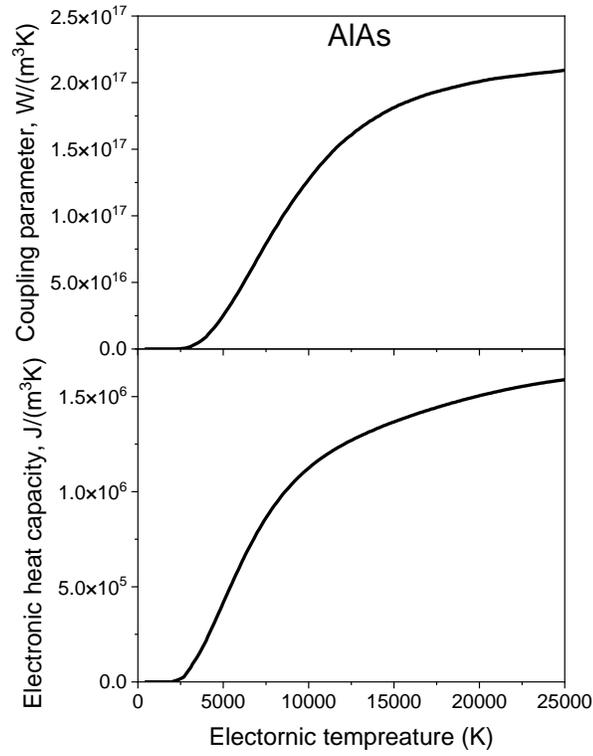

*Figure 7. (Top panel) Electron-phonon coupling, and (bottom panel) electronic heat capacity in AlAs.*

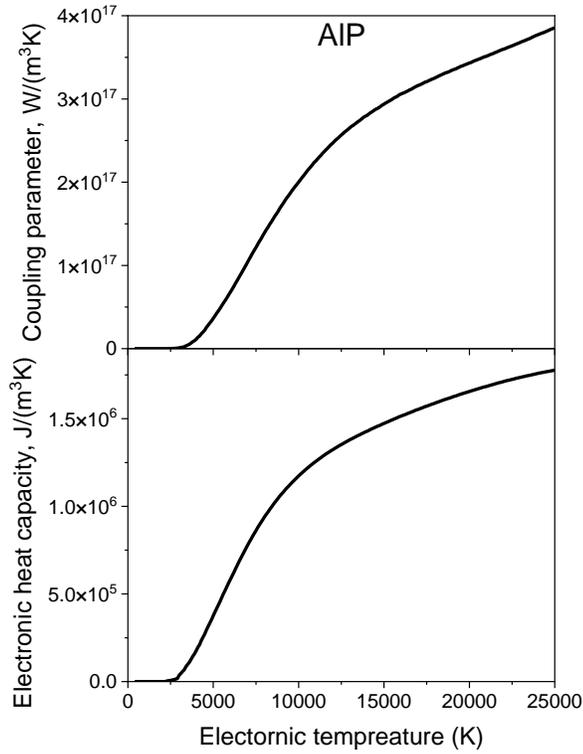

*Figure 8. (Top panel) Electron-phonon coupling, and (bottom panel) electronic heat capacity in AlP.*

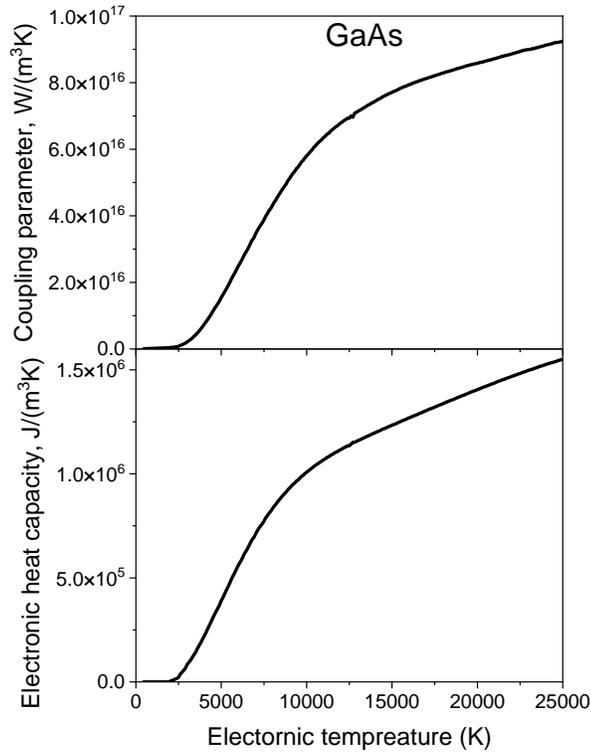

*Figure 9. (Top panel) Electron-phonon coupling, and (bottom panel) electronic heat capacity in GaAs.*

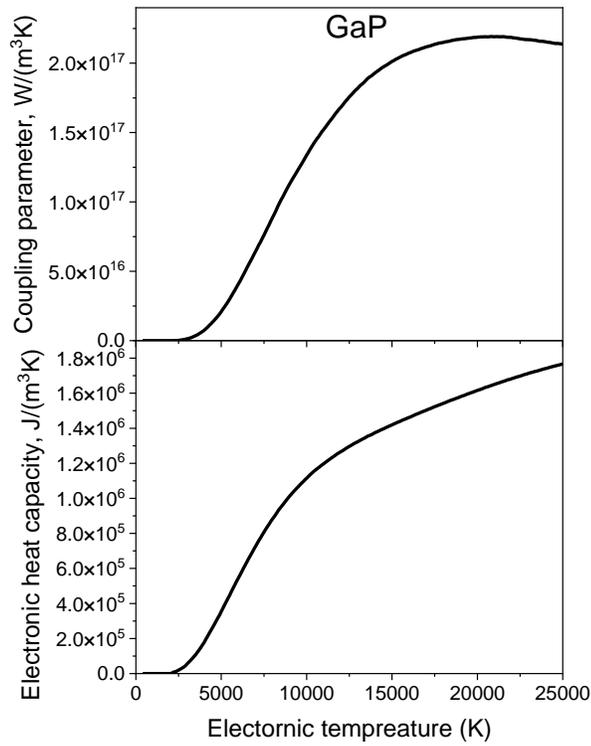

*Figure 10. (Top panel) Electron-phonon coupling, and (bottom panel) electronic heat capacity in GaP.*

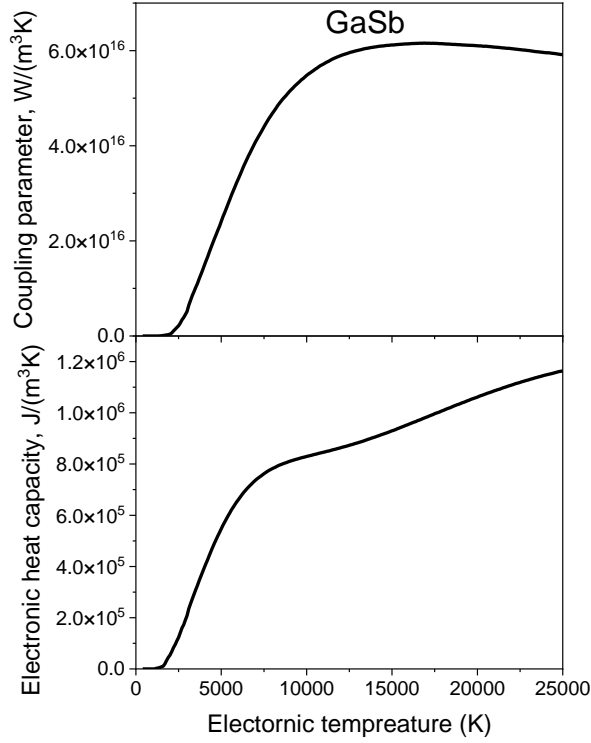

*Figure 11. (Top panel) Electron-phonon coupling, and (bottom panel) electronic heat capacity in GaSb.*

### III.4. Oxides semiconductors

For modeling of oxides semiconductors $Cu_2O$ (384 atoms in the simulation box in cubic Pn3m structure [34]) and $TiO_2$ (216 atoms in the simulation box in rutile structure), matsci-0-3 DFTB parameterization was used [35]; and znorg-0-1 DFTB parameterization for ZnO (384 atoms in hexagonal P6$_3$mc structure [34]) was employed [38].

In all these materials, the coupling parameter is quite high (due to the presence of the light element – oxygen), reaching the values of $G$~$10^{18}$ W/(m$^3$K) at the electronic temperature of $T_e$~25000 K in $TiO_2$. Again, aligning with the overall trend of the decreasing coupling parameter with the atomic mass, a smaller coupling is in $Cu_2O$, and the smallest in ZnO.

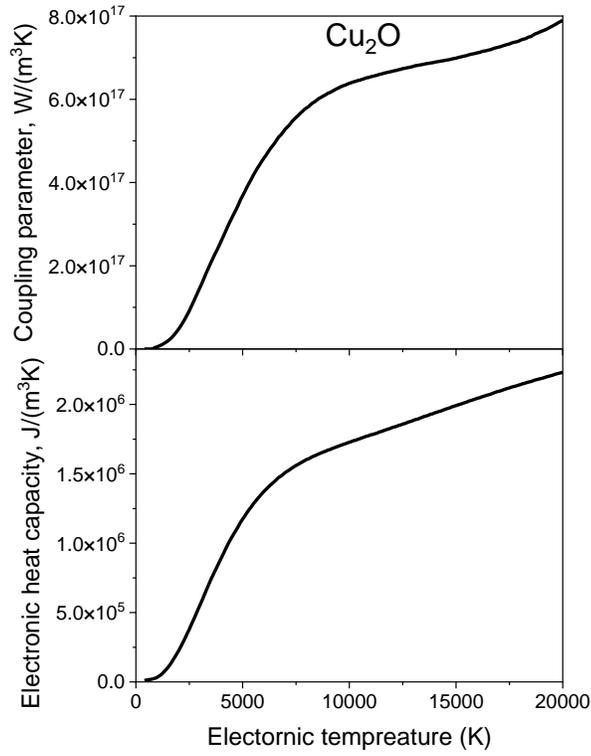

*Figure 12. (Top panel) Electron-phonon coupling, and (bottom panel) electronic heat capacity in Cu₂O.*

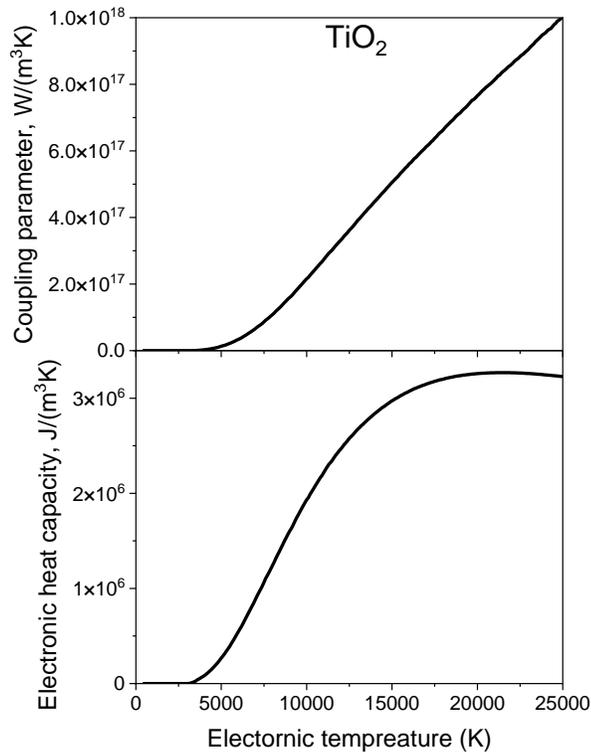

*Figure 13. (Top panel) Electron-phonon coupling, and (bottom panel) electronic heat capacity in TiO₂.*

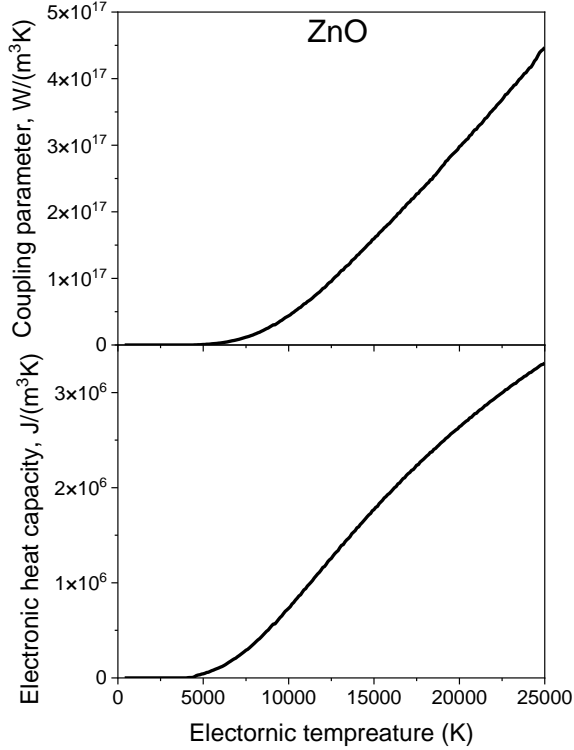

*Figure 14. (Top panel) Electron-phonon coupling, and (bottom panel) electronic heat capacity in ZnO.*

### III.5. Other types of semiconductors

Three other types of semiconductors were also studied: $B_4C$ (270 atoms in trigonal R-3m structure [34]) using matsci-0-3 DFTB parameterization [35]; group II-VI semiconductor ZnS, modeled with znorg-0-1 DFTB parameterization (252 atoms in trigonal $P_3m_1$ structure [34]) [38]; and layered $PbI_2$ (consisting of 192 atoms in hexagonal $P6_3mc$ structure [34]) with the DFTB-based parameterization from Ref. [39] with added ZBL-short-range repulsive potential similar to the method described in Ref. [40] (a detailed description of the parameterization of $PbI_2$ will be published elsewhere).

Made of the lightest elements studied in this work, $B_4C$ demonstrates the highest coupling parameter, $G\sim3.5\times10^{18}$ W/(m$^3$K) at the electronic temperature of $T_e\sim25000$ K. The electron-phonon coupling parameters in $PbI_2$ and ZnS are comparable, both reaching $G\sim1.5\times10^{17}$ W/(m$^3$K) at $T_e\sim20000$ K. Note that $PbI_2$ is an ionic, not covalent, compound, which does not exhibit such pronounced nonthermal effects as bandgap collapse [41], but this point is beyond the scope of the present work.

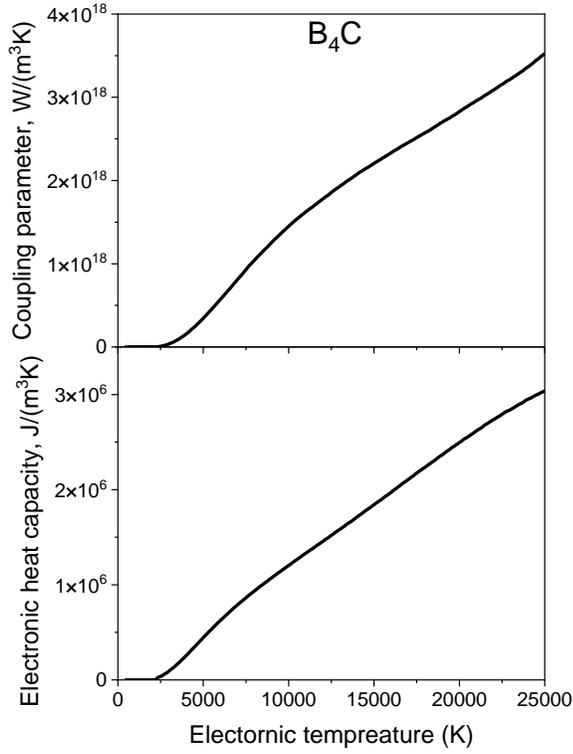

*Figure 15. (Top panel) Electron-phonon coupling, and (bottom panel) electronic heat capacity in $B_4C$.*

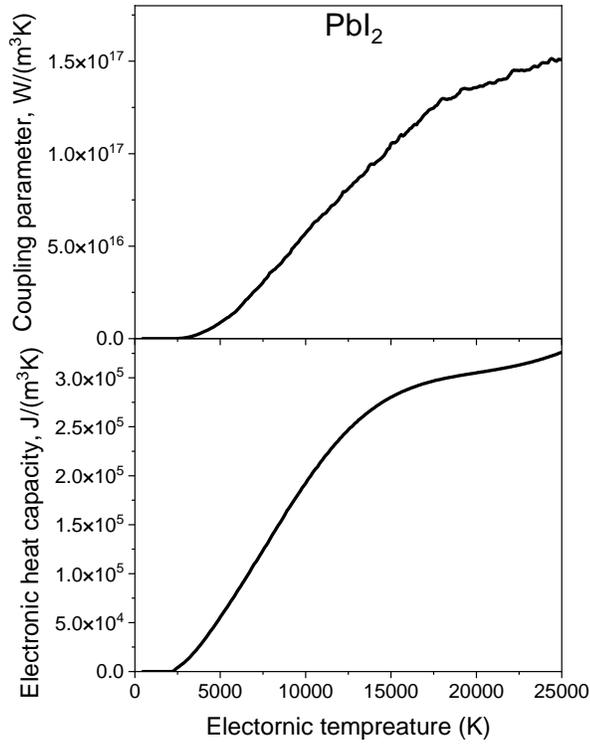

*Figure 16. (Top panel) Electron-phonon coupling, and (bottom panel) electronic heat capacity in $PbI_2$.*

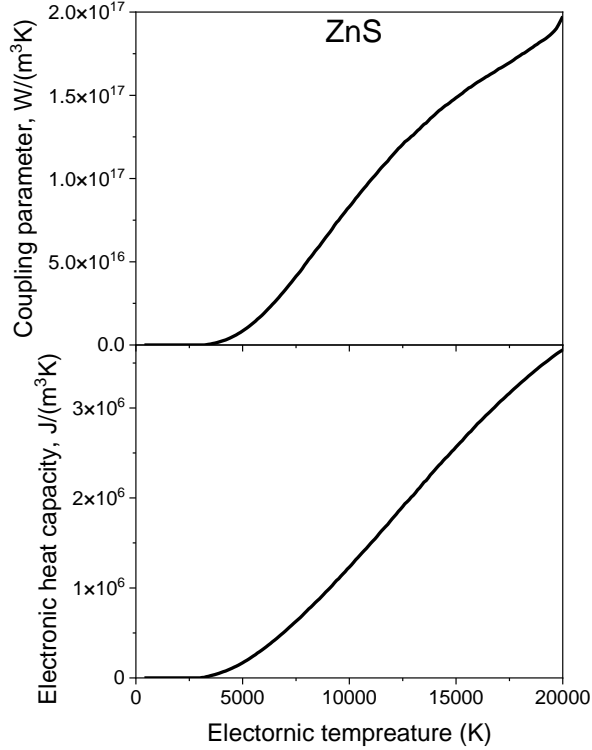

*Figure 17. (Top panel) Electron-phonon coupling, and (bottom panel) electronic heat capacity in ZnS.*

## IV. Conclusions

This work analyzed the electron-ion (electron-phonon) coupling parameters in various semiconductors with the help of the XTANT-3 hybrid model. It is based on the combined Boltzmann equation with tight-binding molecular dynamics. It employs dynamical coupling formalism to evaluate the matrix elements entering the electron-phonon coupling in the nonperturbative regime, used in the Boltzmann collision integral.

It was demonstrated that the coupling parameter is only mildly sensitive to the state of the electronic system at a constant deposited energy: fully non-equilibrium electronic distribution, partially band-resolved thermalized (separate equilibrium in the valence and conduction bands at different temperatures and chemical potentials), and fully equilibrium Fermi-Dirac distributions, all produced the coupling parameters different only within ~35% at high electronic kinetic temperature. In many cases, the difference was even smaller, suggesting that the equilibrium values of the coupling parameter could be used in the modeling of laser irradiation of semiconductors, such as the two-temperature or three-temperature models (and variations or extensions thereof).

The coupling parameters in various semiconductors as functions of the electronic temperatures were calculated up to the electronic temperatures of ~25000 K, which is typically a temperature around the onset of nonthermal melting in covalent materials. All the coupling parameters calculated follow the overall trend of decreasing coupling with the increase of the atomic mass.

## Acknowledgments

Computational resources were supplied by the project "e-Infrastruktura CZ" (e-INFRA LM2018140) provided within the program Projects of Large Research, Development, and Innovations Infrastructures. The author gratefully acknowledges financial support from the Czech Ministry of Education, Youth, and Sports (grants No. LTT17015, LM2018114, and No. EF16_013/0001552).

## Data availability

All the calculated electron-phonon coupling parameters and electronic heat capacities are available online in [42] (together with the previous data for metals from Ref. [20] and 2d-carbon materials from Ref. [29]).

## Appendix. Effect of TB Parameterization

It was previously demonstrated that the tight binding parameterization affects rather strongly the electron-phonon coupling in metals [43]. In semiconductors, the conclusion is the same: the TB parameterization may strongly affect the calculated electronic properties: the coupling parameter and the electronic heat capacity, see an example of Si in Figure 18, especially at high electronic temperatures. That suggests that it is extremely important to validate the calculated parameters in future experiments at elevated electronic temperatures. As of now, unfortunately, there is no experimental data available.

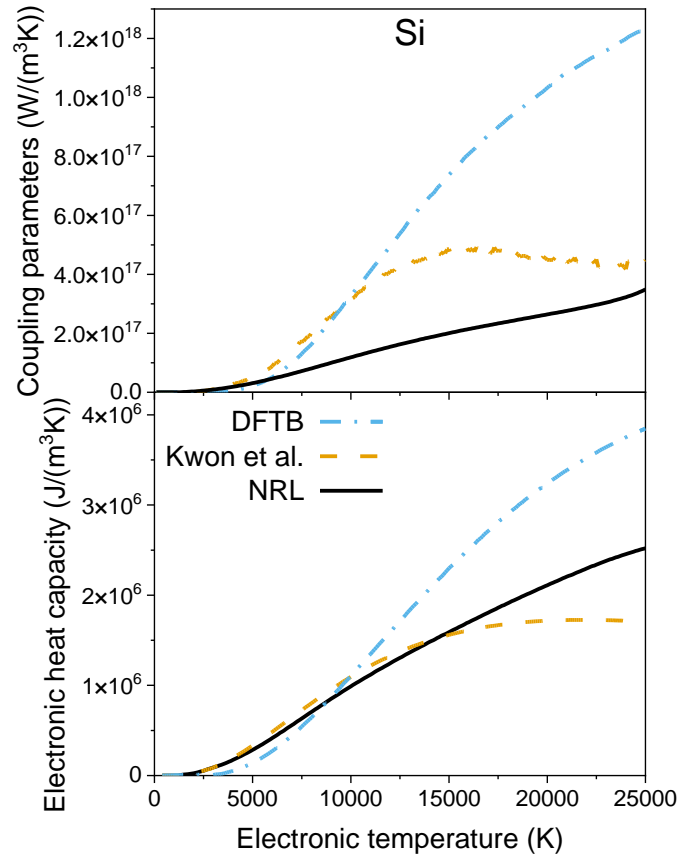

*Figure 18. Comparison of the XTANT-3 calculated coupling parameters (top panel) and electronic heat capacity (bottom panel) in Si for various TB parameterizations: NRL [33], matsci-0-3 DFTB [35], and by Kwon et al. [44],*

## References

[1] W. L. Brown, *Laser Processing of Semiconductors*, **3**, 337 (1983).

[2] A. Elasser and T. P. Chow, *Silicon Carbide Benefits and Advantages for Power Electronics Circuits and Systems*, Proc. IEEE **90**, 969 (2002).

[3] Y. Siegal, E. N. Glezer, L. Huang, and E. Mazur, *Laser-Induced Phase Transitions in Semiconductors*, Annu. Rev. Mater. Sci. **25**, 223 (1995).

[4] P. G. Neudeck, R. S. Okojie, and L. Y. Chen, *High-Temperature Electronics - A Role for Wide Bandgap Semiconductors?*, Proc. IEEE **90**, 1065 (2002).

[5] B. G. Rasheed and M. A. Ibrahem, *Laser Micro/Nano Machining of Silicon*, Micron **140**, 102958 (2021).

[6] J. Wang, F. Fang, H. An, S. Wu, H. Qi, Y. Cai, and G. Guo, *Laser Machining Fundamentals: Micro, Nano, Atomic and Close-to-Atomic Scales*, Int. J. Extrem. Manuf. **5**, 012005 (2023).

[7] F. Rossi and T. Kuhn, *Theory of Ultrafast Phenomena in Photoexcited Semiconductors*, Rev. Mod. Phys. **74**, 895 (2002).

[8] J. Bonse and S. Gräf, *Maxwell Meets Marangoni—A Review of Theories on Laser-Induced Periodic Surface Structures*, Laser Photon. Rev. **14**, 2000215 (2020).

[9] T.-H. Dinh, N. Medvedev, M. Ishino, T. Kitamura, N. Hasegawa, T. Otobe, T. Higashiguchi, K. Sakaue, M. Washio, T. Hatano, A. Kon, Y. Kubota, Y. Inubushi, S. Owada, T. Shibuya, B. Ziaja, and M. Nishikino, *Controlled Strong Excitation of Silicon as a Step towards Processing*

*Materials at Sub-Nanometer Precision*, Commun. Phys. **2**, 150 (2019).

[10] I. M. Lifshits, M. I. Kaganov, and L. V. Tanatarov, *On the Theory of Radiation-Induced Changes in Metals*, J. Nucl. Energy. Part A. React. Sci. **12**, 69 (1960).

[11] S. I. Anisimov, B. L. Kapeliovich, and T. L. Perel-man, *Electron Emission from Metal Surfaces Exposed to Ultrashort Laser Pulses*, J. Exp. Theor. Phys. **39**, 375 (1974).

[12] B. Rethfeld, D. S. Ivanov, M. E. Garcia, and S. I. Anisimov, *Modelling Ultrafast Laser Ablation*, J. Phys. D. Appl. Phys. **50**, 193001 (2017).

[13] A. Kaiser, B. Rethfeld, M. Vicanek, and G. Simon, *Microscopic Processes in Dielectrics under Irradiation by Subpicosecond Laser Pulses*, Phys. Rev. B **61**, 11437 (2000).

[14] N. S. Shcheblanov and T. E. Itina, *Femtosecond Laser Interactions with Dielectric Materials: Insights of a Detailed Modeling of Electronic Excitation and Relaxation Processes*, Appl. Phys. A **110**, 579 (2012).

[15] V. Lipp, B. Rethfeld, M. Garcia, and D. Ivanov, *Solving a System of Differential Equations Containing a Diffusion Equation with Nonlinear Terms on the Example of Laser Heating in Silicon*, Appl. Sci. **10**, 1853 (2020).

[16] A. Rämer, O. Osmani, and B. Rethfeld, *Laser Damage in Silicon: Energy Absorption, Relaxation, and Transport*, J. Appl. Phys. **116**, 053508 (2014).

[17] P. Venkat and T. Otobe, *Three-Temperature Modeling of Laser-Induced Damage Process in Silicon*, Appl. Phys. Express **15**, 041008 (2022).

[18] P. B. Allen, *Theory of Thermal Relaxation of Electrons in Metals*, Phys. Rev. Lett. **59**, 1460 (1987).

[19] M. Toulemonde, C. Dufour, A. Meftah, and E. Paumier, *Transient Thermal Processes in Heavy Ion Irradiation of Crystalline Inorganic Insulators*, Nucl. Instruments Methods Phys. Res. Sect. B Beam Interact. with Mater. Atoms **166–167**, 903 (2000).

[20] N. Medvedev and I. Milov, *Electron-Phonon Coupling in Metals at High Electronic Temperatures*, Phys. Rev. B **102**, 064302 (2020).

[21] N. Medvedev, *Electronic Nonequilibrium Effect in Ultrafast-Laser-Irradiated Solids*, Https://Arxiv.Org/Abs/2302.09098v1 (2023).

[22] N. Medvedev, *XTANT-3: X-Ray-Induced Thermal And Nonthermal Transitions in Matter: Theory, Numerical Details, User Manual*, Http://Arxiv.Org/Abs/2307.03953 (2023).

[23] J. G. Powles, G. Rickayzen, and D. M. Heyes, *Temperatures: Old, New and Middle Aged*, Mol. Phys. **103**, 1361 (2005).

[24] N. Medvedev and I. Milov, *Electron-Phonon Coupling and Nonthermal Effects in Gold Nano-Objects at High Electronic Temperatures*, Materials (Basel). **15**, 4883 (2022).

[25] N. Medvedev and A. E. Volkov, *Nonthermal Acceleration of Atoms as a Mechanism of Fast Lattice Heating in Ion Tracks*, J. Appl. Phys. **131**, 225903 (2022).

[26] C. W. Siders, A. Cavalleri, K. Sokolowski-Tinten, C. Tóth, T. Guo, M. Kammler, M. H. von Hoegen, K. R. Wilson, D. von der Linde, and C. P. J. Barty, *Detection of Nonthermal Melting by Ultrafast X-Ray Diffraction*, Science **286**, 1340 (1999).

[27] A. Rousse, C. Rischel, S. Fourmaux, I. Uschmann, S. Sebban, G. Grillon, P. Balcou, E. Förster, J. P. Geindre, P. Audebert, J. C. Gauthier, and D. Hulin, *Non-Thermal Melting in Semiconductors Measured at Femtosecond Resolution.*, Nature **410**, 65 (2001).

[28] G. J. Martyna and M. E. Tuckerman, *Symplectic Reversible Integrators: Predictor–Corrector Methods*, J. Chem. Phys. **102**, 8071 (1995).

[29] N. Medvedev, I. Milov, and B. Ziaja, *Structural Stability and Electron-phonon Coupling in Two-dimensional Carbon Allotropes at High Electronic and Atomic Temperatures*, Carbon Trends **5**, 100121 (2021).

[30] B. Rethfeld, A. Kaiser, M. Vicanek, and G. Simon, *Ultrafast Dynamics of Nonequilibrium Electrons in Metals under Femtosecond Laser Irradiation*, Phys. Rev. B **65**, 214303 (2002).

[31] N. Medvedev, F. Akhmetov, R. A. Rymzhanov, R. Voronkov, and A. E. Volkov, *Modeling

*Time-Resolved Kinetics in Solids Induced by Extreme Electronic Excitation*, Adv. Theory Simulations **5**, 2200091 (2022).

[32] M. J. Mehl and D. A. Papaconstantopoulos, *NRL Transferable Tight-Binding Parameters Periodic Table: Http://Esd.Cos.Gmu.Edu/Tb/Tbp.Html*.

[33] D. A. Papaconstantopoulos and M. J. Mehl, *The Slater Koster Tight-Binding Method: A Computationally Efficient and Accurate Approach*, J. Phys. Condens. Matter **15**, R413 (2003).

[34] *Materials Project*, https://materialsproject.org/.

[35] J. Frenzel, A. F. Oliveira, N. Jardillier, T. Heine, and G. Seifert, Semi-Relativistic, Self-Consistent Charge Slater-Koster Tables for Density-Functional Based Tight-Binding (DFTB) for Materials Science Simulations., 2009.

[36] N. Medvedev, Z. Li, and B. Ziaja, *Thermal and Nonthermal Melting of Silicon under Femtosecond X-Ray Irradiation*, Phys. Rev. B **91**, 54113 (2015).

[37] C. Molteni, L. Colombo, and L. Miglio, *Tight-Binding Molecular Dynamics in Liquid III-V Compounds. I. Potential Generation*, J. Phys. Condens. Matter **6**, 5243 (1994).

[38] N. H. Moreira, G. Dolgonos, B. Aradi, A. L. da Rosa, and T. Frauenheim, *Toward an Accurate Density-Functional Tight-Binding Description of Zinc-Containing Compounds*, J. Chem. Theory Comput. **5**, 605 (2009).

[39] *DFTB Parameterization without Short-Range Repulsion*, https://github.com/by-student-2017/Slater-Koster-parameters-no-repulsion_v1.

[40] W. Ding, H. He, and B. Pan, *Development of a Tight-Binding Model for Cu and Its Application to a Cu-Heat-Sink under Irradiation*, J. Mater. Sci. **50**, 5684 (2015).

[41] R. A. Voronkov, N. Medvedev, and A. E. Volkov, *Dependence of Nonthermal Metallization Kinetics on Bond Ionicity of Compounds*, Sci. Rep. **10**, 13070 (2020).

[42] N. Medvedev, *Electron-Phonon Coupling and Related Parameters Calculated with XTANT-3*, https://github.com/N-Medvedev/XTANT-3_coupling_data.

[43] F. Akhmetov, N. Medvedev, I. Makhotkin, M. Ackermann, and I. Milov, *Effect of Atomic-Temperature Dependence of the Electron–Phonon Coupling in Two-Temperature Model*, Materials (Basel). **15**, (2022).

[44] I. Kwon, R. Biswas, C. Wang, K. Ho, and C. Soukoulis, *Transferable Tight-Binding Models for Silicon*, Phys. Rev. B **49**, 7242 (1994).